# The first CCD photometric studies of the member eclipsing binary ZTFJ015003.88+534734.1 in the newly discovered young open cluster UBC 188

Y. H. M. Hendy[1*], I. Zead[1], A. E. Abdelaziz[1], A. Takey[1]

[1]National Research Institute of Astronomy and Geophysics (NRIAG), 1142 Helwan, Cairo, Egypt

*Corresponding author.
E-mail address: yasserhendy@nriag.sci.eg (Y. H. M. Hendy).

## Abstract

We present the first CCD observations of an eclipsing binary, ZTFJ015003.88+534734.1, which is a member in the open star cluster UBC 188. The observations were taken by the 1.88 m telescope at the Kottamia Astronomical Observatory (KAO) in SDSS *griz* bands. The latest version of the Wilson-Devinney (W-D) code was employed for photometric analysis and light curve modeling of the eclipsing binary. The results indicate that the binary system is in an over-contact configuration. The mass of the primary star (M1) is determined to be 1.293 $M_\odot$, and the mass of the secondary star (M2) is directly derived from the system's estimated mass ratio (q= M2/M1) as 0.340 times the solar mass ($M_\odot$). We investigated the color-magnitude diagram and the membership probability of the open cluster UBC 188 using the Gaia DR3 data. We determined the membership probability of the eclipsing binary ZTFJ015003.88+534734.1 using the pyUPMASK algorithm and found that its membership probability is one.

**Keywords:** Variable stars - Eclipsing binary stars - W UrsaeMajoris variable stars- Open star clusters.

## 1. Introduction

Eclipsing binary systems belong to a broad category of variable stars, where the variability arises from the mutual eclipses of two stars as they orbit a common center of gravity. The variation occurs when one star is obscured by the other from the perspective of an observer. This event leads to a reduction in the brightness of the binary system. W UMa-type contact binaries are generally composed of two late-type stars that share a common convective envelope. This type of detached binaries system is short period binaries. The two components of the system usually merge into a single rapid rotation star or blue straggler due to magnetic braking tidal friction, and angular momentum loss (AML) (Bradstreet & Guinan 1994; Qian et al. 2006; Li et al. 2019). The companion may play an essential role during the process by removing angular momentum and energy exchanges (Eggleton & Kisseleva-Eggleton 2006; Fabrycky & Tremaine 2007; Qian et al. 2014, 2018).

The light curves of the contact binaries have nearly equal depths of the two eclipsing minima and the continuous light variability, revealing that the two components of the contact binaries have similar temperatures. The observed light curve serves as a representation of the system's processes and their changes over time (Prsa 2018). Although many efforts have been made in more than five decades to comprehend their formation, evolution, structure, and ultimate destiny, no theoretical model can clearly and superiorly interpret their observed properties (Lucy 1968; Qian 2003; Stepien 2006, 2011; Eggleton 2012). Furthermore, the short-period cutoff, the O'Connell effect, the minimum mass ratio, and the thermal relaxation oscillation theory are still controversial (O'Connell 1951; Flannery 1976; Lucy 1976; Lucy & Wilson 1979; Rucinski 1992; Rasio 1995; Qian 2001; Li & Zhang 2006; Qian et al. 2014; Zhou et al. 2016; Li et al. 2019; Caton et al. 2019; Li et al. 2021).

The ZTFJ015003.88+534734.1 or ATO J027.5161+53.7927 is an eclipsing binary candidate (Heinze 2018). The ZTFJ015003.88+534734.1 is a W UMA

eclipsing binary system as defined by the ZTF catalog[1] through the IRSA Catalog, which is a completely automated, wide-field survey designed for the systematic investigation of the optical transient sky. The coordinates of the eclipsing binary are ra J (2000): 01 50 03.88 and dec J (2000): +53 47 34.1, taken from the ZTF catalog. This paper presents the first time series of the *griz* CCD observations and data analysis of the ZTFJ015003.88+534734.1.

Castro-Ginard et al. (2020) discovered the open cluster UBC 188. They determined the sky position of the UBC 188 in ra= 27.446±0.19 deg and dec=53.924±0.15 deg. Dias et al. (2021) measured the photometric parameters of the UBC 188 of log(age/yr)=7.22±0.12, extinction $A_v$=0.70±0.08 mag, and the photometric distance 2577±161 pc. Hunt and Reffert (2023) calculated the photometric parameters of the UBC 188 of log(age/yr)=7.56, extinction $A_v$=0.59 mag, and the photometric distance was 2618 pc.

The study of eclipsing binaries within star clusters serves as an excellent assessment for theories related to the structure and evolution of binary systems. ZTFJ015003.88+534734.1 is an unstudied eclipsing binary star located at 0.11 deg from the center of the newly discovered open cluster UBC 188. Cantat-Gaudin et al. (2020) measured that the radius containing half the members $r_{50}$ of the UBC 188 equals 0.142 deg.

The structure of the paper is outlined in the following manner: observations and data reduction are presented in section 2. Light curve analysis and evolution status of the ZTFJ015003.88+534734.1 are given in section 3 and section 4, respectively. Star cluster membership and color-magnitude diagram of the UBC 188 are presented in section 5. Finally, the conclusion is presented in section 6.

---

[1] https://irsa.ipac.caltech.edu/Missions/ztf.html

## 2. Observations and Data Reduction

The data used in the present work were carried out using the Kottamia Faint Imaging Spectro-Polarimeter (KFISP) system attached to the Cassegrain focus of the 1.88-m telescope at the Kottamia Astronomical Observatory (KAO), Egypt (Azzam et al. 2010). KFISP has a 2K × 2K CCD camera (E2V 42-40); more details about KFISP are given by Azzam et al. (2022).

We take our observations in the Sloan filters *griz*. These filters are used to obtain the time series of the W UMA system ZTFJ015003.88+534734.1. The system was observed over two nights, Table 1 (1-2, November 2022). Figure 1 shows the field of the binary system ZTFJ015003.88+534734.1, which is taken using the KAO, 1.88-m telescope. The observation log of the binary system is given in Table 1.

All times are corrected to HJD, the reduction of the observed frames. The bias, the flat correction, and aperture photometry were performed using the software package C-Munipack (Motl 2007). Then, the differential magnitudes between the variable and the comparison star were derived.

Table 1. KAO observation log of our target.

| Filter | Number of Images | Exposure Time(s) |
|---|---|---|
| g | 57 | 180 |
| r | 60 | 180 |
| i | 56 | 180 |
| z | 60 | 180 |

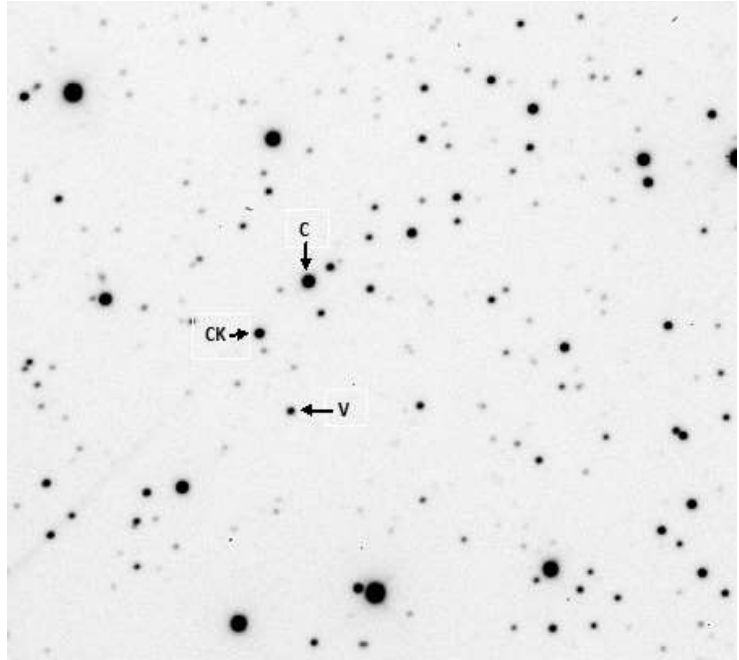

Figure 1. An image of the field containing the binary system ZTFJ015003.88+534734.1 taken in the SDSS -*r* band by KFISP attached to the 1.88-m telescope at the KAO. The size of the image is 8.2′x8.2′. The variable, comparison, and check stars are identified as V, C, and CK, respectively. The direction of the North is up, and the East is left.

The times of primary and secondary minima of ZTFJ015003.88+534734.1 were obtained for the *griz* bands and presented in Table 2. The following equation represents the new ephemeris performed for the system by using Kwee and van Woerden's method (1956).

$$\text{HJD (MinI)} = 2459886.4878(\pm 0.0009) + 0.3821^{d} \times E$$

The E is the number of integer cycles. This ephemeris calculates the phases and draws the light curves in the *griz* bands, as shown in Figure 2.

Table 2. The primary and secondary minima of the ZTFJ015003.88+534734.1 were derived in our analysis.

| Filter | MinI | MinII |
|---|---|---|
| g | 2459886.487375±0.0007 | 2459886.297829±0.00120 |
| r | 2459886.486863±0.0096 | 2459886.289415±0.00030 |
| i | 2459886.487790±0.0009 | 2459886.291298±0.00002 |
| z | 2459886.487790±0.0080 | 2459886.291298±0.00040 |

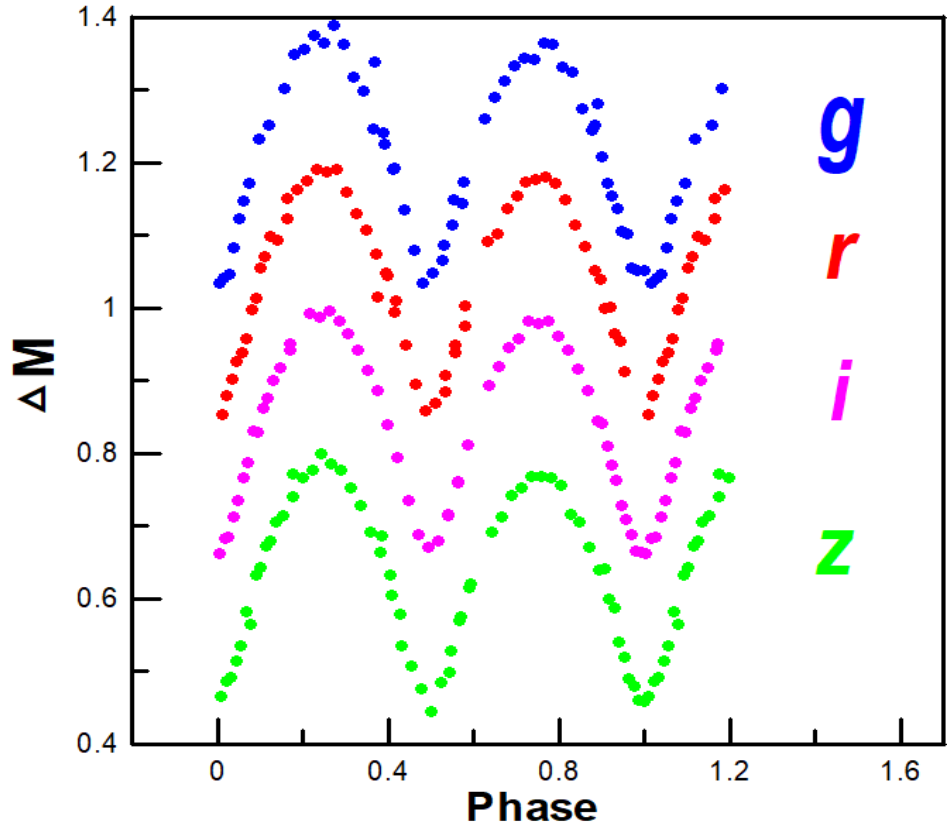

Figure 2. The observed phase diagram of ZTFJ015003.88+534734.1 in the *griz* bands was taken by our telescope.

## 3. Light Curve Analysis of the ZTFJ015003.88+534734.1

The observed light curves of the ZTFJ015003.88+534734.1 system in *griz* bands were subjected to photometric analysis using the Wilson-Devinney (W-D) program (Nelson 2009; Wilson and Devinney 1971). In this analysis, we considered gravity darkening and bolometric albedo exponents suitable for the convective envelopes of late spectral type stars (Teff < 7500 K), as indicated by Abdelaziz et al. (2020). We adopted $g_1=g_2=0.32$ (Lucy 1967) and $A_1=A_2=0.5$ (Rucinski 1969). Limb darkening values were adopted using Van Hamme (1993). We obtained the effective temperature of the primary star ($T_1$) as 6330 K from the spectroscopic survey of the LAMOST DR9 database portal[2].

So far, there are no published mass results based on spectroscopic or photometric data. To establish the initial mass ratio, we employed a q-search approach. As outlined by Terrell and Wilson (2005), there exists a connection between the q value and the system's inclination, and in cases involving partial eclipses—such as our system, which is an overcontact system undergoing partial eclipses—the uncertainty of q tends to increase (Abdelaziz et al. 2022). In our analysis, the determination of q reached convergence with an indeterminate value within an error margin of 2.28%. Figure 3 illustrates the achieved convergence with q at 0.263±0.006.

The refined parameters encompass the orbital inclination (i), mean temperature of the secondary star ($T_2$), potential of both stars ($\Omega=\Omega_1=\Omega_2$), mass ratio (q), and luminosity of the primary star ($L_1$). We conducted the analysis utilizing Mode 3 (over contact mode, not thermal contact) within the Wilson-Devinney (W-D) program. Examination of the light curves indicates that the primary star surpasses the secondary one in temperature, with an approximate difference of 60 K between the two stars. The parameters for the accepted solution are given in Table 3, while Figure 4 lists the optimal fit of the model to the observed light curve. Also, Figure 4 shows light-curve asymmetries, with the magnitudes at phase 0.25 being different from those at phase 0.75. This is called 'the O'Connell effect', which is assumed to be caused by spot activity. In order to model the asymmetric light curves, the spots model of (W- D) was used. A cool spot on either of the two components was applied by Li et al. (2019). The spot is located on the primary component of the system. It is also shown in Figure 5 at phase 0.75. From the iteration, the temperature factor is 0.3 and it is a small factor. This maight be because it is a small and cold spot. The spectral types of the two stars in the ZTFJ015003.88+534734.1 system are F7 and F8, as reported by Covey et al. (2007), in accordance with the estimated temperature.

---
[2] http://www.lamost.org/dr9/

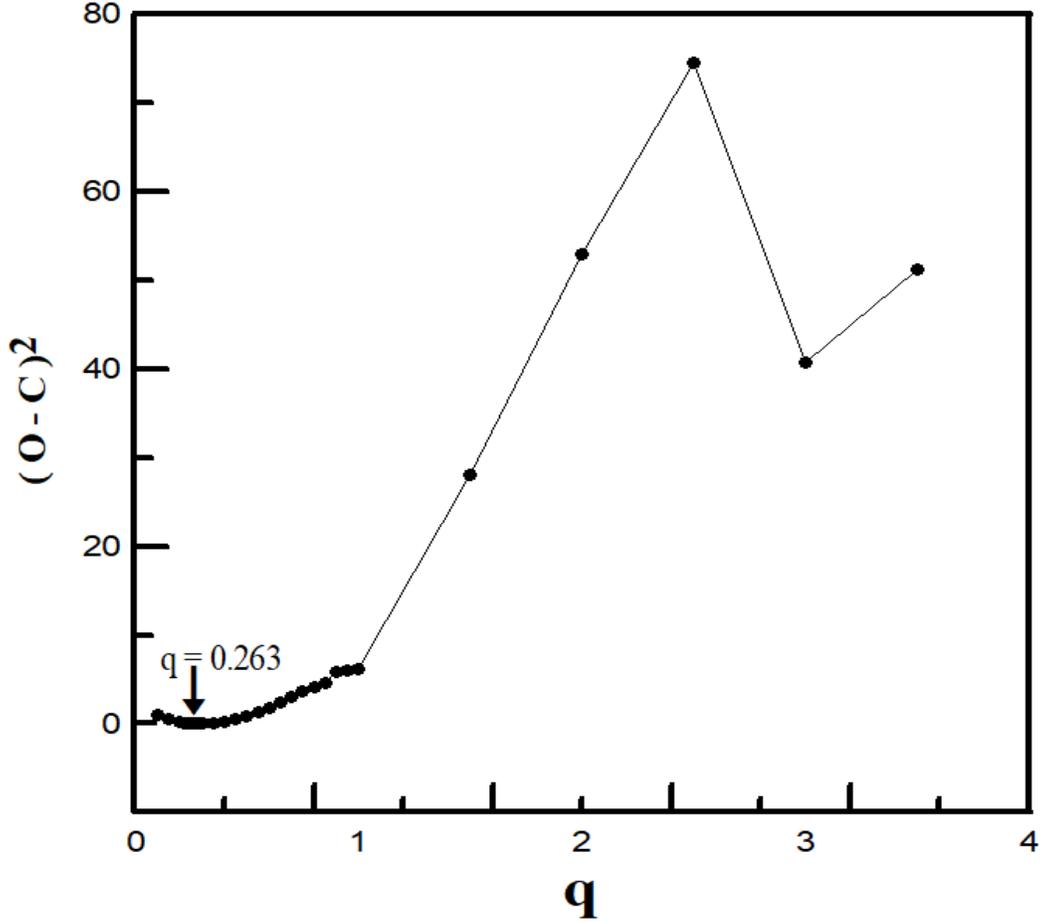

Figure 3. q - Σ (O – C)$^2$ relation for the system ZTFJ015003.88+534734.1

Figure 5 illustrates the geometric arrangement of the binary system. Employing the empirical relation proposed by Harmanec (1988), we computed the absolute physical parameters for both stars in the study system. The primary star's mass, M1, is determined to be 1.293 times the solar mass (M$_\odot$), while the secondary star's mass, M2, is directly derived from the estimated mass ratio of the system (q = M2 /M1) as 0.340 times the solar mass (M$_\odot$). The radii of both stars, R1(R$_\odot$) and R2(R$_\odot$), along with their bolometric magnitudes, M1bol and M2bol, were computed and are presented in Table 4.

Table 3. The orbital solution for ZTFJ015003.88+534734.1 across the *griz* bands.

| Parameter | g | r | i | z |
|---|---|---|---|---|
| λ | 4680 Å | 6170 Å | 7480 Å | 8930 Å |
| T1(K) | 6330 ± 137 | 6330 ± 137 | 6330 ± 137 | 6330 ± 137 |
| T2 (K) | 6230 ± 34 | 6300 ± 32 | 6270 ± 13 | 6270 ± 25 |
| q | 0.263±0.017 | 0.263±0.010 | 0.263±0.006 | 0.263±0.011 |
| Ω1=Ω2 | 2.3490±0.034 | 2.3395±0.026 | 2.3540±0.012 | 2.3506±0.026 |
| g1=g2 | 0.32 | 0.32 | 0.32 | 0.32 |
| A1=A2 | 0.5 | 0.5 | 0.5 | 0.5 |
| X1=X2 | 0.644±0.078 | 0.644±0.067 | 0.644±0.030 | 0.644±0.055 |
| Y1=Y2 | 0.231 | 0.231 | 0.231 | 0.231 |
| i(°) | 76.4±1.9 | 74±1.23 | 74.7±0.58 | 75.07±1.14 |
| r pole1 | 0.47364 | 0.47574 | 0.47254 | 0.47329 |
| r side1 | 0.51316 | 0.51611 | 0.51162 | 0.51267 |
| r back1 | 0.53992 | 0.54362 | 0.53801 | 0.53931 |
| r pole2 | 0.25972 | 0.26204 | 0.25852 | 0.25934 |
| r side2 | 0.27144 | 0.27421 | 0.27001 | 0.27098 |
| r back2 | 0.31101 | 0.31612 | 0.30842 | 0.31017 |
| L1/(L1+L2) | 0.794±0.03 | 0.774±0.03 | 0.784±0.01 | 0.784±0.02 |
| ∑(o−c)2 | 0.015 | 0.011 | 0.003 | 0.008 |
| fill-out factor | 20.5 % | 26.3 % | 17.5 % | 19.5 % |
| co-latitude | 90 | 90 | 90 | 90 |
| co-longitude | 90 | 90 | 90 | 90 |
| Radius | 5 | 5 | 5 | 5 |
| Temperature factor | 0.3 | 0.3 | 0.3 | 0.3 |

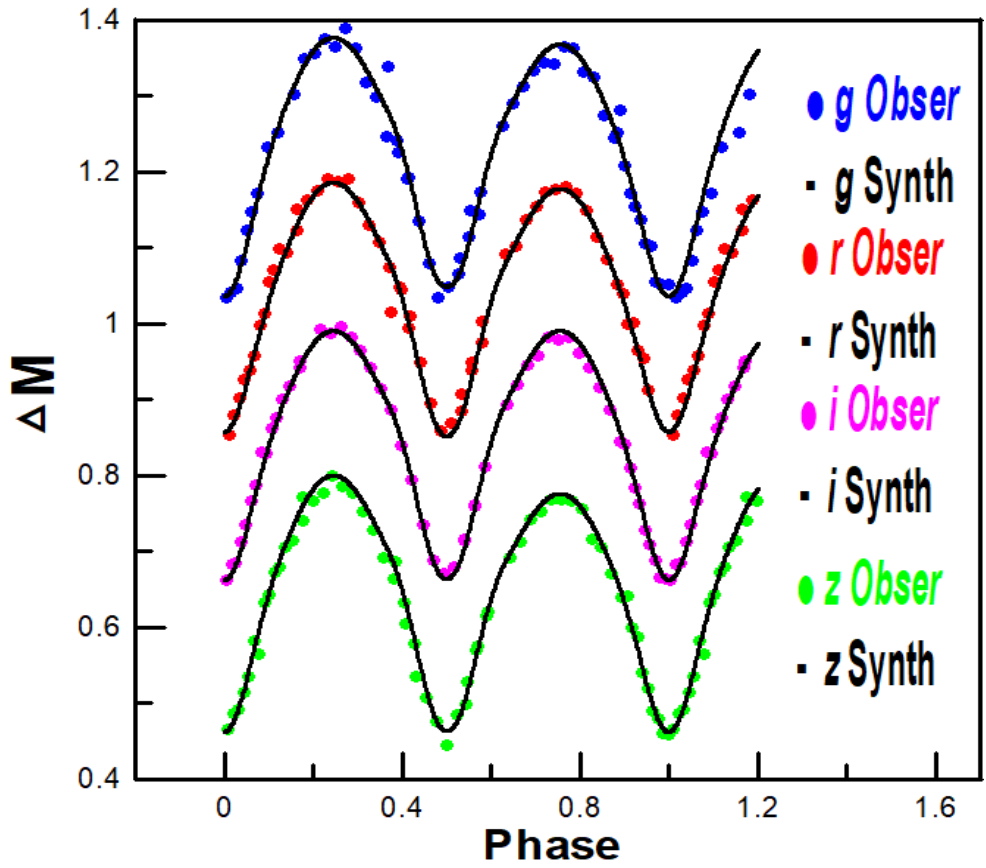

Figure 4. Light curves in the *griz* of the system ZTFJ015003.88+534734.1, with the synthetic model fitted with a spot to the data represented by the solid line.

Table 4. The absolute physical characteristics of ZTFJ015003.88+534734.1 are derived from our analysis.

| Element | M (M$_\odot$) | R (R$_\odot$) | T (T$_\odot$) | L (L$_\odot$) | M$_{bol}$ | Sp.type | Log g |
|---|---|---|---|---|---|---|---|
| Primary | 1.293 | 1.379 | 1.096 | 2.739 | 3.600 | F7 | 4.270 |
| Secondary | 0.340 | 1.361 | 1.085 | 2.568 | 3.670 | F8 | 4.275 |

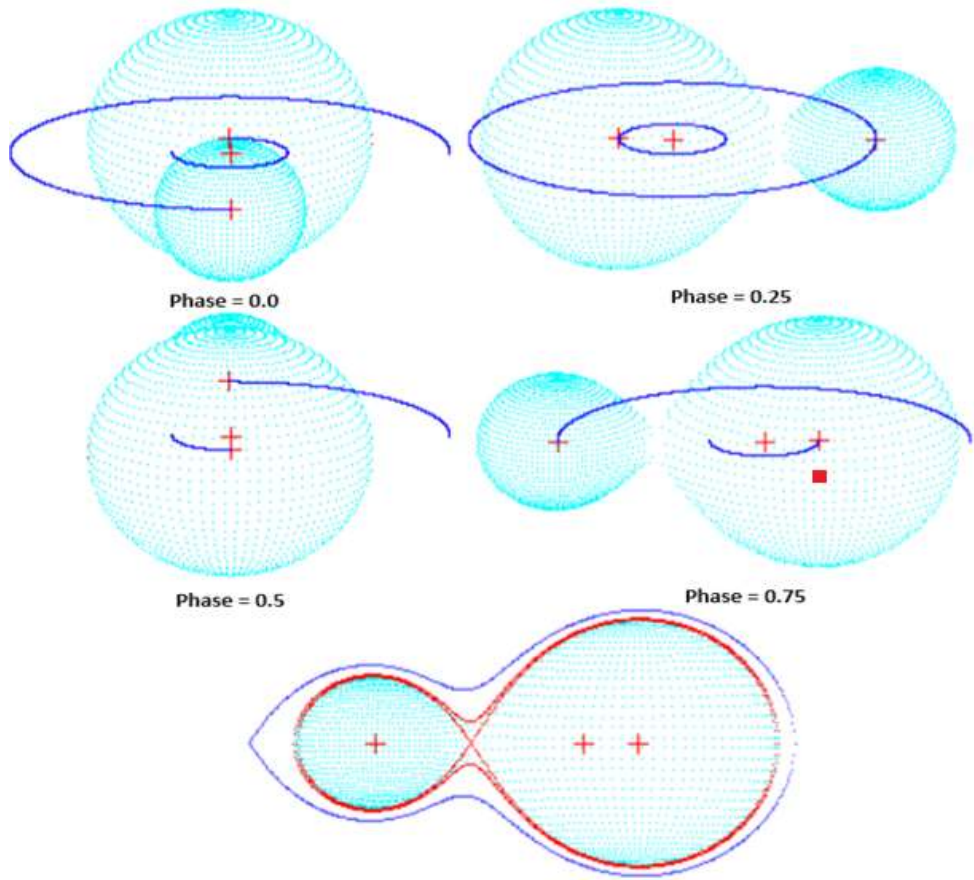

Figure 5. The geometric arrangement at various phases and the configuration of ZTFJ015003.88+534734.1. The spot is shown at phase 0.75 marked by red square.

## 4. Evolution Status of the ZTFJ015003.88+534734.1

Utilizing the derived physical parameters presented in Table 4, we examined the current evolutionary status of the system. Figure 6 displays the system's components on the Mass-Luminosity (M-L) relation (left panel), Mass-Radius (M-R) relation (middle panel), and the Mass-Temperature (M-T) relation (right panel). Additionally, the figure includes the evolutionary tracks computed by Mowlavi et al. (2012) for both the Zero Age Main Sequence (ZAMS) and Terminal Age Main Sequence (TAMS) with a metallicity of z=0.014. Also, the right panel of Figure 6 presents the positions of the system components on the empirical Mass-Teff relation for intermediate-mass stars by Malkov (2007). As evident from Figure 6, in the left and middle panels, the primary star aligns with the ZAMS track, while the secondary star, in both relations, has evolved beyond the TAMS track. The right panel shows that the primary star is perfectly located on Mass-Teff for intermediate-mass stars derived by Malkov (2007), while the secondary star is completely off from the relation.

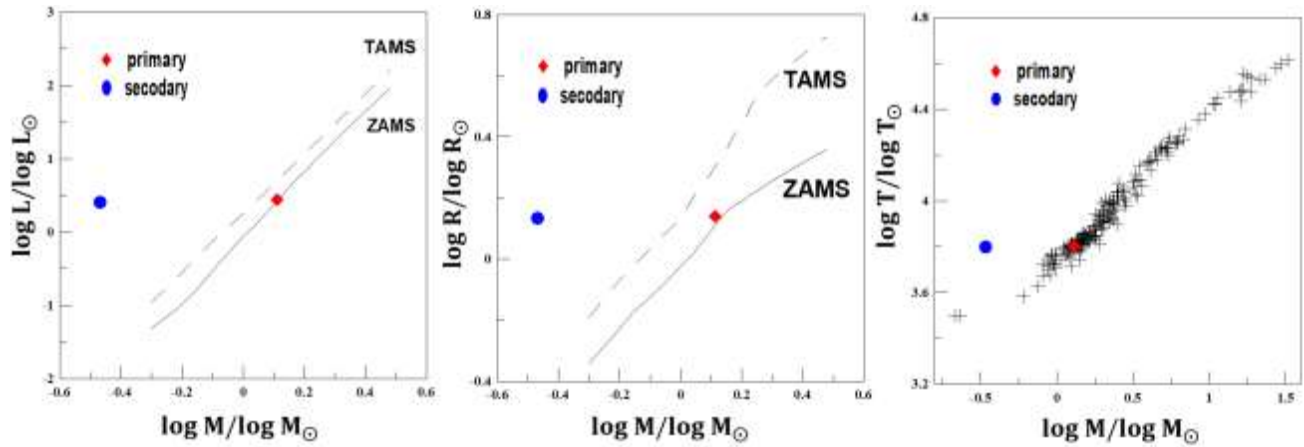

Figure 6. The placement of both stars within the system on the Mass-Luminosity (left), Mass-Radius (middle) relation according to Mowlavi et al. (2012), and the empirical Mass-Teff relation (right) for intermediate-mass stars based on Malkov (2007).

## 5. Star Cluster Membership and Color-Magnitude Diagram of the UBC 188

To determine cluster member stars, we download all surrounding stars within 0.3 degrees from the center of the open cluster UBC 188 using the Gaia DR3[3]. The center of the UBC 188 is ra=27.466 and dec=53.900 (Cantat-Gaudin et al. 2020). The search radius is enough to identify the cluster membership.

We used the clustering technique pyUPMASK to obtain membership probability (Pera et al. 2021), as we have used it in Casado and Hendy (2024). To get a good astrometric quality, we use Re-normalised Unit Weight Error (RUWE)< 1.4. We selected the most probable members with a membership probability> 0.5. This yielded 97 member stars of the UBC 188. Figure 7 presents the vector point diagram of the proper motions in ra and dec, the sky position in ra and dec, and the comoving stars of the members and field stars. Figure 8 shows the parallax-G diagram of the members and field stars.

---

[3] https://gea.esac.esa.int/archive/

We used the PARSEC theoretical isochrones of Bressan et al. (2012) with the solar metallicity value Z=0.0152. We calculated the fundamental parameters (age, distance modulus, and extinction) using the ASTECA algorithm (Perren et al. 2015), as we have used in Casado and Hendy (2023). The best fit of the Color-Magnitude Diagram (CMD) of the cluster is verified with the isochrones of log (age/yr)=7.34±0.07, distance modulus $(m-M)_0$=12.09±0.07 mag and extinction $A_v$=0.56±0.04 mag (Figure 9). Cantat-Gaudin et al. (2020), Dias et al. (2021), and Hunt and Reffert (2023) determined the ranges of the log(age/yr) of the UBC 188 are 7.42, 7.22±0.12, and 7.56 respectively. Our measurement of the log(age/yr) is comparable to the measurements of Cantat-Gaudin et al. (2020) and Dias et al. (2021).

We utilized $A_G$=0.84 $A_v$ and E(BP-RP)=0.45 $A_v$, employing absorption ratios from various photometric systems at different wavelengths (Cardelli et al. 1989, O'Donnell 1994). Consequently, we derived an extinction value of $A_G$=0.47±0.05 mag and a reddening value of E(BP-RP)=0.25±0.03 mag.

The photometric distance based on the distance modulus of the UBC 188 is 2618±119 pc. Using the inverted corrected parallax (i.e., mean parallax+0.017, Lindegren et al. 2021), we obtained the astrometric distance= 2632±649 pc. Our measurements of photometric distance and astrometric distance are in good agreement. Hunt and Reffert (2023) determined log(age/yr)=7.56, extinction $A_v$=0.59 mag, and the photometric distance 2618 pc. Our results of the extinction and the photometric distance are consistent with their results.

The parallax of the ZTFJ015003.88+534734.1 system is 0.366 mas. Using the inverted corrected parallax, we obtained the astrometric distance as 2611±349 pc. The membership probability of this binary system is one. Our measurement of its membership probability is also one determined by Cantat-Gaudin et al. (2020). The photometric distance of the open cluster UBC 188 and the astrometric distance of the eclipsing binary are consistent with each other.

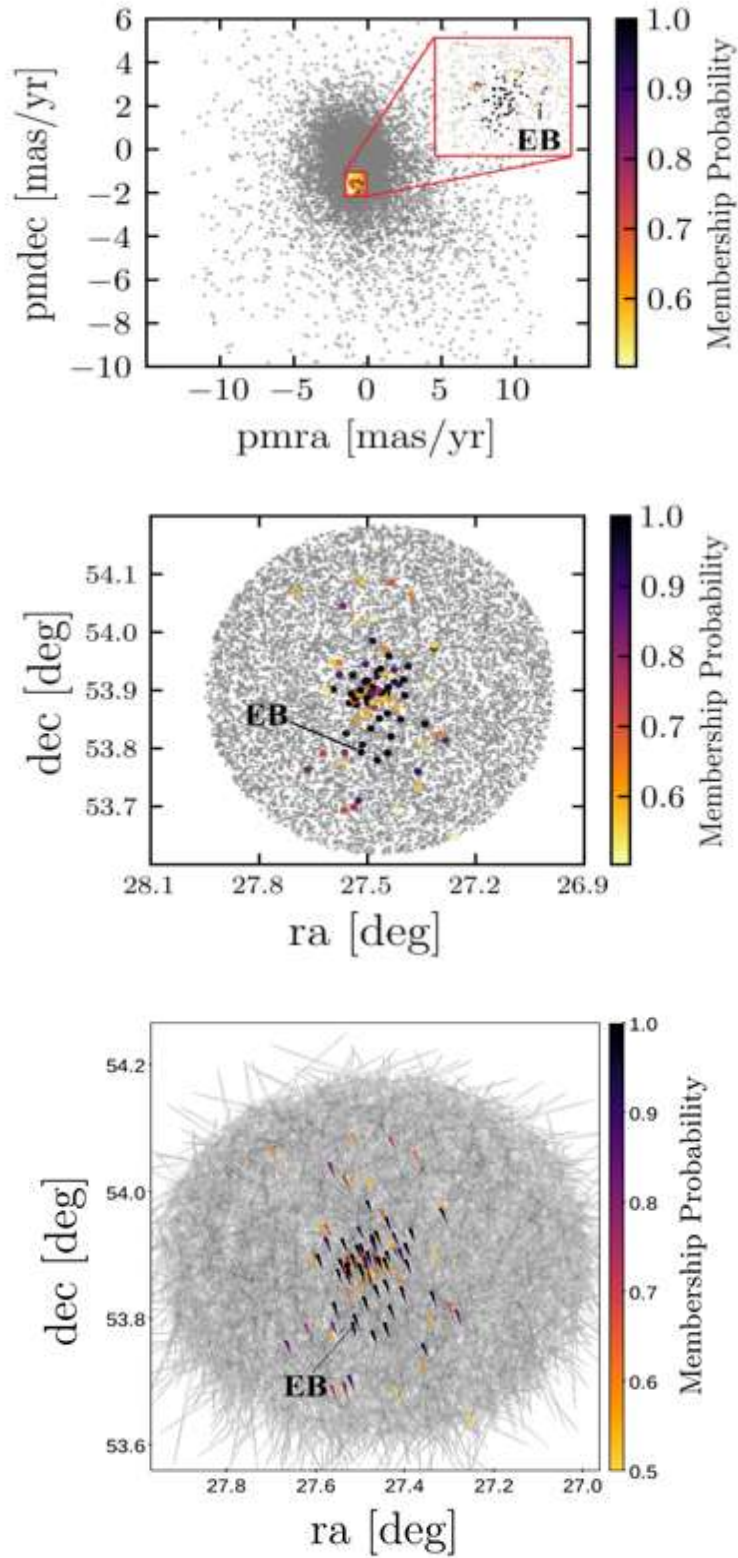

Figure 7. The vector point diagram is in the upper panel, the sky position in the middle panel, and the comoving stars of the open cluster UBC 188 in the bottom panel. The eclipsing binary is marked in all panels.

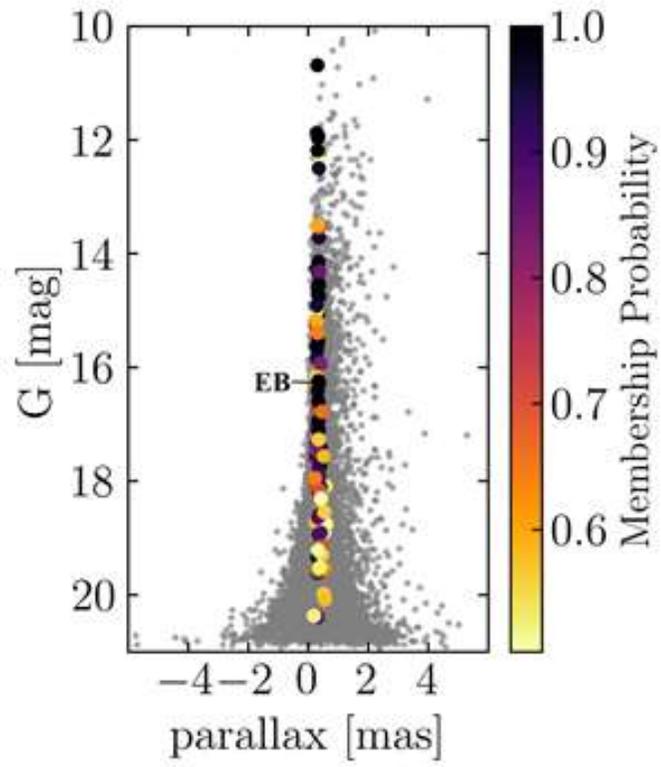

Figure 8. The parallax-G diagram for the eclipsing binary and the open cluster UBC 188. Its binary member is also indicated.

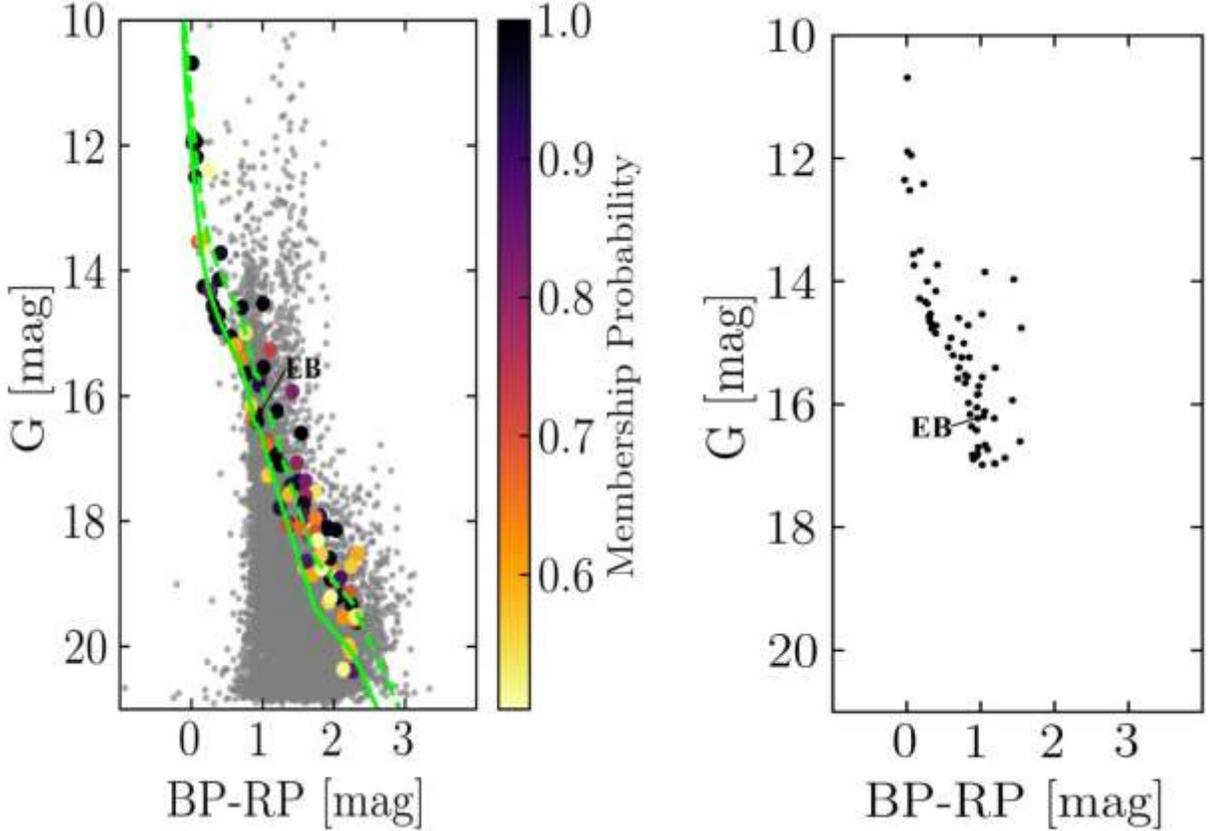

Figure 9. The color-magnitude diagram (CMD) from our investigation is in the left panel. The solid lines represent PARSEC isochrones fitted to our data, while the dashed lines depict the same isochrones vertically shifted by -0.75 mag (representing the locus of unresolved binaries with equal mass components). The right panel shows the CMD from the study conducted by Cantat-Gaudin et al. (2020). In both panels, the binary system is indicated by EB.

We calculated the astrophysical parameters of the open cluster UBC 188 and found 97 members with a membership probability>0.5. We noted that the number of cluster members determined in our study, 97 stars, is larger than the number of 69 stars identified by Cantat-Gaudin et al. (2020). This is because we have used a fainter limiting magnitude (G < 21) in Gaia DR3. The mean astrometric measurements of proper motions and parallax are pmra=-0.81±0.18 mas/yr, pmdec= -1.62±0.17mas/yr, and parallax= 0.36±0.08 mas. The mean sky position in ra and dec are 27.48±0.08 deg and 53.88±0.08 deg, respectively. All the measured parameters of the cluster are listed in our work in Table 5.

Table 5. The physical parameters of the open cluster UBC 188 are measured in our study. Other published parameters are also given.

| Parameter | Our study | Cantat-Gaudin et al. (2020) |
|---|---|---|
| ra (deg) | 27.48±0.08 | 27.47 |
| dec (deg) | 53.88±0.08 | 53.90 |
| pmra (mas/yr) | -0.81±0.18 | -0.87±0.25 |
| pmdec (mas/yr) | -1.62±0.17 | -1.56±0.19 |
| parallax (mas) | 0.36±0.08 | 0.34±0.03 |
| Members (stars) | 97 | 69 |
| log (age/yr) | 7.34±0.07 | 7.42 |
| $A_v$ (mag) | 0.56±0.04 | 0.33 |
| $A_G$ (mag) | 0.47±0.05 | - |
| $(m-M)_0$ (mag) | 12.09±0.07 | 12.29 |
| E(BP-RP) (mag) | 0.25±0.03 | - |
| photometric distance (pc) | 2618±119 | 2872 |
| astrometric distance (pc) | 2632±649 | - |

## 6. Conclusion

We observed the time series of the eclipsing binary ZTFJ015003.88+534734.1 for the first time in the Sloan bands *griz*. Using a 1.88-m telescope at the Kottamia Astronomical Observatory (KAO) in Egypt. This eclipsing binary is a member in the young open star cluster UBC 188, which is also studied in our work based on Gaia DR3 data. Our main results are summarized as follows:

- We analyzed the light curves of *griz* bands of the system ZTFJ015003.88+534734.1 using the (W–D) code to obtain the geometric and photometric parameters of the system.
- The ZTFJ015003.88+534734.1 system is an overcontact binary with a fill-out factor =17.5% and a low mass ratio=0.263.
- The binary system is classified as A-subtype. The primary star is the massive and hotter one. The spectral types of the primary and secondary stars are F7 and F8, respectively.
- The positions of both stars of ZTFJ015003.88+534734.1 on both Mass–Radius and Mass–Luminosity relations revealed that the primary star is a main sequence star. At the same time, the secondary is an evolved star.
- The ZTFJ015003.88+534734.1 is at 0.11 deg from the center of the newly discovered open cluster UBC 188.
- We calculated the physical parameters of the open cluster UBC 188, as given in Table 5, using color-magnitude diagrams of the Gaia DR3. We determined the age, reddening, and photometric distance as log(age/yr) = 7.34±0.07, E(BP-RP)=0.25±0.03 mag and 2618±119 pc, respectively.
- Using the high precision of the astrometric of Gaia DR3 proper motions, we found that the eclipsing binary ZTFJ015003.88+534734.1 is a member star with a membership probability equal to one and has an astrometric distance of 2611±349 pc.

## Acknowledgments
This paper is based upon work supported by Science, Technology & Innovation Funding Authority (STDF) under grant number STDF 45779. We thank the reviewer for his/her valuable comments that helped to improve the paper. This paper used data from the European Space Agency (ESA) mission Gaia (https://www.cosmos.esa.int/web/gaia/dr3) and the Guoshoujing Telescope (the Large Sky Area Multi-Object Fiber Spectroscopic Telescope LAMOST). The LAMOST is operated and managed by the National Astronomical Observatories, Chinese Academy of Sciences. This work made use of the SIMBAD database (https://simbad.unistra.fr/simbad/) and the VizieR catalog access tool, operating at the CDS, Strasbourg, France (https://vizier.cds.unistra.fr/viz-bin/VizieR), and of NASA Astrophysics Data System (https://ui.adsabs.harvard.edu/).